\documentclass[prb,reprint,noeprint,superscriptaddress]{revtex4-1}
\usepackage{graphicx}
\usepackage{subfigure}
\usepackage{nicefrac}
\usepackage{color}
\usepackage{amsmath}
\usepackage{amssymb}
\usepackage{amsfonts}
\usepackage{amsthm}
\usepackage{mathrsfs}
\usepackage{wasysym}
\usepackage{bm}
\usepackage{placeins}
\usepackage[bookmarks=false]{hyperref}
\usepackage[usenames,dvipsnames]{xcolor}
\usepackage{physics}
\usepackage{chemformula}
\usepackage{mathtools}

\usepackage{pifont}
\usepackage[dvipsnames]{xcolor}
\colorlet{commentcolor}{green!10!orange!90!}

\begin{document}

\title{Fermi surface reconstruction by a charge-density-wave with finite correlation length}

\author{Yuval Gannot}
\affiliation{Department of Physics, Stanford University, Stanford, California 94305, USA.}
\author{Brad J. Ramshaw}
\affiliation{Laboratory of Atomic and Solid State Physics, Cornell University, Ithaca, NY 14853, USA}
\author{Steven A. Kivelson}
\affiliation{Department of Physics, Stanford University, Stanford, California 94305, USA.}

\begin{abstract}
Even a small amplitude charge-density-wave (CDW) can reconstruct a Fermi surface, giving rise to new quantum oscillation frequencies. Here, we investigate quantum oscillations when the CDW has a finite correlation length $\xi$ -- a case relevant to the hole-doped cuprates. By considering the Berry phase induced by a spatially varying CDW phase, we derive an effective Dingle factor that depends exponentially on the ratio of the cyclotron orbit radius, $R_c$, to $\xi$. In the context of YBCO, we conclude that the values of $\xi$ reported to date for bidirectional CDW order are, prima facie, too short to account for the observed Fermi surface reconstruction; on the other hand, the values of $\xi$ for the unidirectional CDW are just long enough.

 \end{abstract}   
\date{\today}

\maketitle

\section{Introduction}
Charge density wave (CDW) order is a common feature of the cuprate superconductors, and is generally believed to be responsible for the Fermi surface reconstruction apparent in quantum oscillation (QO) experiments
\cite{Sebastian_Proust_2015,Proust_Taillefer_2019}. In YBCO, two kinds of CDW are observed in x-ray scattering: a bidirectional CDW and a field induced unidirectional CDW \cite{Comin_Damascelli_2016}. 
Most commonly, the Fermi surface reconstruction is attributed to the bidirectional CDW \cite{Harrison_Sebastian_2011, Laliberte_et_al_2018}. However, there are also proposals for reconstruction by the field induced unidirectional CDW \cite{Yao_Lee_Kivelson_2011, Millis_Norman_2007} -- consistent with the fact that the unidirectional CDW appears above $\approx 15 \text{ T}$, whereas QOs are observed only above  $\approx 18\text{ T}$ \cite{Maharaj_et_al_2016}.

An important observation is that the correlation length of the bidirectional CDW is rather short: $\xi_{2Q} \approx 100 \text{ \AA}$ \cite{Chang_et_al_2012}, whereas the cyclotron radius at the lowest fields for which QOs are observed is $R_c \approx 400 \text{ \AA}$. Since QOs are expected to be strongly damped when $\xi \lesssim R_c$, it is unclear whether the observed signal is consistent with reconstruction by the bidirectional CDW. 

On the other hand, the correlation length of the unidirectional CDW is longer: $\xi_{1Q} \approx 200 \text{ \AA}$ \cite{Chang_et_al_2016} at fields relevant for QOs, which could more easily account for experimental observations. A quantitative understanding of QOs in disordered CDWs may therefore help distinguish between the two proposed reconstruction scenarios.

With this experimental motivation in mind, we undertake a theoretical investigation of Fermi surface reconstruction by a CDW with finite correlation length. Our primary  result is an 
expression for the Dingle factor $R_{D}(p)$ which suppresses the amplitude of the $p$th QO harmonic. For the first harmonic and for the reconstruction scenarios relevant to the cuprates, 
\begin{equation}
R_D(p=1) = e^{-B_D/B}; \ \   B_D = \frac{ 2n \hbar k_F}{e\xi}. \label{eq:B_D_intro}
\end{equation}
Here $\xi$ is the relevant CDW correlation length, $n = 1$ for unidirectional order and $n=2$ for bidirectional order, and $2k_F$ is defined to be the distance between points on the Fermi surface at which the CDW Bragg scatters the electron. Eq. ~\ref{eq:B_D_intro} is a lower bound on the experimentally observed Dingle field $B^{\text{exp}}_D$, as it neglects all disorder besides the finite CDW correlation length.

Combining this result with the measured correlation lengths $\xi_{1Q}$ and $\xi_{2Q}$, we predict
\begin{equation}
B^{\text{exp}}_D \gtrsim
\begin{cases}
90 \text{ T} & \text{unidirectional order} \\
340 \text{ T} & \text{bidirectional order}
\end{cases}
\end{equation}
in \ch{YBa2Cu3O_{$6.59$}}, whereas from QO measurements we find $B^{\text{exp}}_D \approx 110 \text{ T}$. That is, the lower bound set by the bidirectional CDW is violated, while the lower bound set by the unidirectional CDW is just satisfied. Given that the QO frequency evolves smoothly with hole doping, this observation is not easy to reconcile with reconstruction by the bidirectional CDW. On the other hand, reconstruction by the unidirectional CDW is marginally consistent.

The effect of CDW phase disorder on the semiclassical spectrum can be expressed in terms of a contribution to a Berry phase each time an electron Bragg scatters off the CDW.  This formulation leads to a remarkable real-space structure to the local density of states in the case of a locally commensurate CDW punctuated by well-separated, sharp discommensurations (DCs). 
For a commensurability $m$ CDW, the Landau level spectrum is shifted by $ \pm \hbar\omega_c/m$  in a region of width $R_c$ about the DC -- something which should be directly observable in scanning tunneling spectroscopy.  In terms of the spectrum of QOs, this has the unusual consequence that while most of the harmonics are suppressed by the same sort of Dingle factor already discussed, if disordering of the CDW  is caused entirely by randomly spaced DCs, the $p = m$ harmonic (and all multiples of it) are not affected at all. To verify this  result, we have reproduced it by exact solution of an explicit lattice-scale model. 

Returning to YBCO, this raises the possibility that the dominant $540 \text{ T}$ QO frequency
observed in experiment
might actually be the sixth harmonic of a $90 \text{ T}$ fundamental. We make no serious assertion that this major re-interpretation of the data is correct -- but it is an interesting possibility naturally suggested by our results, and which could resolve other experimental discrepancies.

The remainder of this paper is organized as follows.
We 
introduce a model Hamiltonian in Sec.~\ref{sec:model}, which we use throughout to illustrate our argument, and as a basis for numerics. In Sec.~\ref{sec:semiclassical_no_cdw}, we briefly review the semiclassical theory of QOs in conventional metals. 
We introduce a heuristic ``scattering picture" to derive an expression for an extra phase $\gamma$ 
that enters the semiclassical quantization condition
in Sec.~\ref{sec:semiclassical_scattering}.
The resulting expression is generalized and evaluated for Fermi surface reconstructions relevant to the cuprates in Sec.~\ref{sec:general_reconstruction} and ~\ref{sec:cuprate_reconstruction}, respectively. In Sec.~\ref{sec:berry}, we 
show that $\gamma$ may also be obtained as 
a Berry phase. In Sec.~\ref{sec:dingle} we derive the Dingle factor and Dingle field, including a discussion of higher harmonics for a random DC array. These theoretical results are confirmed by numeric simulations in Sec.~\ref{sec:numerics}. 
Finally, we apply these results to experiments in \ch{YBa2Cu3O_{$y$}} in Sec.~\ref{sec:experiments}.

\section{Model \label{sec:model}}
Throughout this paper, we describe weak CDW order by an effective Hamiltonian $H = H_0 + U$, where $H_0$ describes the underlying crystal and $U$ is the CDW potential. Here, we introduce a specific model used to illustrate our results. We take
\begin{equation} 
H_0 =  -t \sum_{\ev{ \bm{r}',\bm{r}} } \pqty{ c^{\dagger}_{\bm{r}'} c_{\bm{r}}  +   \text{h.c.}}, 
\end{equation}
where $c^{\dagger}_{\bm{r}}$ creates an electron at position $\bm{r} = (x,y)$ on a square lattice and $\ev{\bm{r}',\bm{r}}$ denotes nearest neighbors. 
$H_0$ can be diagonalized as
\begin{equation}
H_0 = \sum_{\bm{k}} \mathcal{E}_0(\bm{k}) c^{\dagger}_{\bm{k}} c_{\bm{k}} \\
\end{equation}
where $c^{\dagger}_{\bm{k}}$ creates an electron in the Bloch state with crystal momentum $\bm{k}$ and 
\begin{equation}
\mathcal{E}_0(\bm{k}) = -2t 
\left[\cos(ak_x) + \cos(ak_y) \right],
\end{equation}
where $a$ is the lattice constant. 

For $E_F<0$, this yields a
roughly circular electron-like Fermi surface centered at $\bm{k} = 0$. For the CDW potential, we take
\begin{equation}
U = 2 V \sum_{\bm{r}} \cos[ \bm{Q} \vdot\bm{r} + \phi(\bm{r})] c^{\dagger}_{\bm{r}} c_{\bm{r}}
\label{cdw}
\end{equation}
where $V>0$ and $\phi(\bm{r})$ is the local phase of the CDW. The assumed CDW is ``weak'' in the sense that $V/E_F \ll 1$.

\section{Review of semiclassical analysis in the absence of  a CDW \label{sec:semiclassical_no_cdw}
}
We first consider the the problem in the absence of a CDW, i.e. $V=0$.
 Assuming the band under consideration has no Berry curvature,
the equations of motion (EOM) for the mean position $\bm{r}$ (now treated as a continuous variable) and gauge invariant crystal momentum $\bm{k}$ of a wave packet are \cite{Ashcroft_Mermin_1976}
\begin{align}
\dot{\bm{r}} &= \frac{1}{\hbar} \grad_{\bm{k}} \mathcal{E}_0(\bm{k}) \label{eq:EOM_1} \\
\hbar \dot{\bm{k}} &= - e \dot{\bm{r}} \cp \bm{B}. \label{eq:EOM_2}
\end{align}
Combining the above,
\begin{equation}
\dot{\bm{k}} =  \frac{e}{\hbar^2} \bm{B} \cp \grad_{\bm{k}} \mathcal{E}_0(\bm{k}), \label{eq:EOM_reduced}
\end{equation}
so the $k$-space orbit coincides with the Fermi surface. 
Since we are considering a problem with a closed Fermi surface, the wave packet executes periodic cyclotron motion in $k$-space. In real space
\begin{equation}
\dot{r}_a  = \frac{\hbar}{eB} \epsilon_{ab} \dot{k}_b  \label{eq:EOM_rotate}
\end{equation}
where $\epsilon$ is the Levi-Cevita symbol and the sum over $b=x,$ $y$ is implicit.
Therefore, to switch between $k$-space and real space we simply rotate and re-scale the trajectory. 

There is an infinite family of cyclotron orbits, each labeled by its time-independent guiding center 
\begin{equation} 
{R}_a  = {r}_a - \frac{\hbar}{eB} \epsilon_{ab} {k}_b.  \label{eq:guiding}
\end{equation} 
Periodic cyclotron motion gives rise to a discrete quantum energy spectrum, determined by the semiclassical quantization condition
\begin{equation}
S(E_n) = 2\pi \pqty{ n+ \frac{1}{2}  },
\end{equation}
where the action 
\begin{equation}
S(E) = \oint \pqty{ \bm{k} -\frac{e}{\hbar}\bm{A}(\bm{r}) } \vdot d \bm{r}
\end{equation}
can loosely be thought of as the phase picked up by a wave packet over the course of a cyclotron period. Using the EOM, it is straightforward to obtain
\begin{equation}
S(E) =  \frac{\hbar A(E)}{eB} \label{eq:action=area}
\end{equation}
where $A(E)$ is the area of the Fermi surface.  Thus,
\begin{equation}
 \frac{\hbar A(E_n)}{eB} = 2\pi \pqty{ n+ \frac{1}{2}  }. \label{eq:QO_quantization}
\end{equation}

If the Fermi energy is fixed, Landau levels cross $E_F$ periodically in $1/B$. This happens with frequency 
\begin{equation}
F = \frac{\hbar A(E_F)}{2\pi e} \label{eq:QO_freq}.
\end{equation}
As most properties of a metal depend on the density of states at $E_F$, this gives rise to conventional QOs. 

\section{Semiclassical analysis with a CDW: scattering picture \label{sec:semiclassical_scattering}}
Here, we use a heuristic ``scattering picture'' to understand dynamics and quantization in the presence of a weak CDW, which may be disordered. This discussion is closely related to work by Pippard on QOs in the presence of lattice dislocations \cite{Pippard_1965}.

\subsection{Dynamics}

\begin{figure}
\includegraphics[width=6 cm]{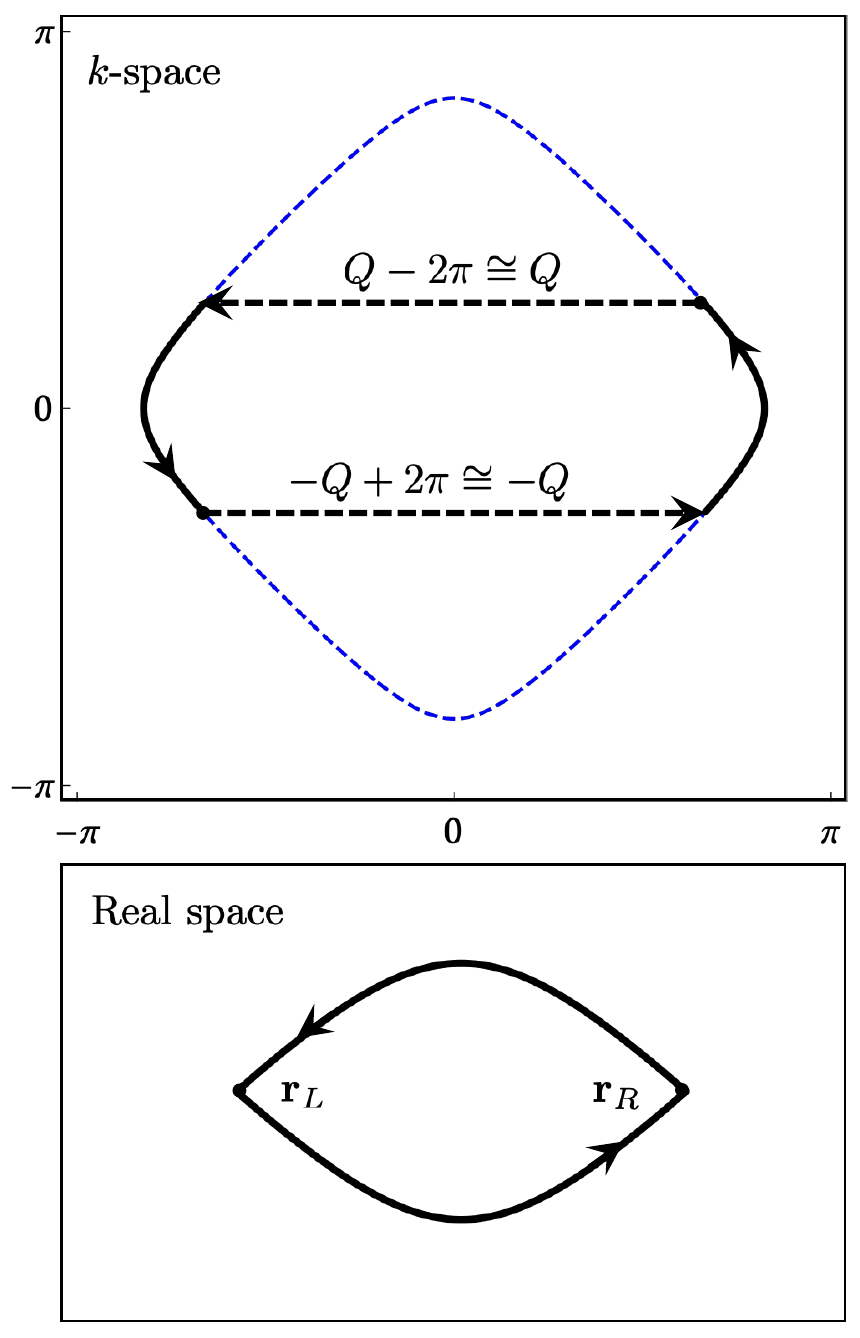}
\caption{\label{fig:model_hamiltonian_reconstruction} 
Reconstruction by a $Q = (1/3)(2\pi/a)$ CDW in the model Hamiltonian. Top: $k$-space cyclotron orbit, visualized in terms of scattering across the unreconstructed Fermi surface. For clarity, each scattering process is folded back into the first Brillouin zone. Bottom: corresponding real-space orbit. }
\end{figure}

In a \emph{weak} CDW, a wave packet evolves along the Fermi surface of $H_0$ essentially as in the absence of the CDW except at discrete ``scattering points'' defined by the nesting condition $\mathcal{E}_0(\bm{k}) = \mathcal{E}_0(\bm{k} \pm \bm{Q})$. Assuming the wave packet is fully scattered from momentum $\bm{k}$ into $\bm{k} \pm \bm{Q}$ (i.e., no magnetic breakdown), the following simplified picture 
applies:
$\bm{k}$ follows the Fermi surface (of $H_0$) until it hits a scattering point, then jumps by $\pm \bm{Q}$ across the Fermi surface, then follows the Fermi surface until the next scattering point, and so on. This is illustrated in Fig.~\ref{fig:model_hamiltonian_reconstruction}. Stitching together the 
segments of the orbit we obtain a closed $k$-space figure which we identify as the reconstructed Fermi surface; when rotated and re-scaled it gives the real space orbit, also indicated in Fig.~\ref{fig:model_hamiltonian_reconstruction}.

Above, we implicitly assumed that the wave packet can scatter only through momenta $\pm \bm{Q}$. This is true if we restrict our attention to first order processes. Higher order processes can scatter the wave packet through arbitrary $ n\bm{Q}$, but the scattering rate will be suppressed relative to first order by powers of $V/t$. Therefore, we expect a broad range of magnetic fields where the the probability of magnetic breakdown is small for first order scattering, but nearly equal to one for higher order scattering. As shown in Appendix~\ref{app:mb}, the appropriate range is 
\begin{equation}
\pqty {\frac{V}{E_F} }^2 \pqty{\frac{V}{v_F}}^2 \ll \frac{eB}{\hbar} \ll \pqty{\frac{V}{v_F}}^2
\label{eq:field_regime}
\end{equation}
where $v_F$ is a characteristic Fermi velocity of the underlying band structure.

Since only first order scattering enters the dynamics, there is no distinction between commensurate and incommensurate CDWs. Moreover, since the CDW is so weak that it only affects the wave packet at the scattering points, the only effect of slightly phase-disordering the CDW is a possible small displacement of the scattering points. A random array of sharp DCs is special in this regard: the semiclassical dynamics are entirely unaffected, except for rare orbits whose scattering points intersect DCs.

\subsection{Quantization}
To quantize this motion, we need to compute the total phase, $S(E)$, picked  up by a wave packet 
of energy $E$ as it executes a single closed orbit:  
\begin{equation}
S(E) = S_F(E) + \theta_R + \theta_L
\end{equation}
where $S_F$ is the action associated with evolution along the Fermi surface, and  $\theta_R$, $\theta_L$, are the phase shifts suffered at the right and left scattering points on the Fermi surface. We denote the corresponding real-space points by $\bm{r}_{R}$, $\bm{r}_L$, as in Fig.~\ref{fig:model_hamiltonian_reconstruction}.

Let us ignore the possible displacements mentioned above; this is justified in Appendix~\ref{app:action_perturbation}. Then $S_F$ is unaffected by disorder. However, $\theta_{R/L}$ depend on the CDW phase at the corresponding scattering points, $\phi(\bm{r}_{R/L})$.

\begin{figure}
\includegraphics[width=6 cm]{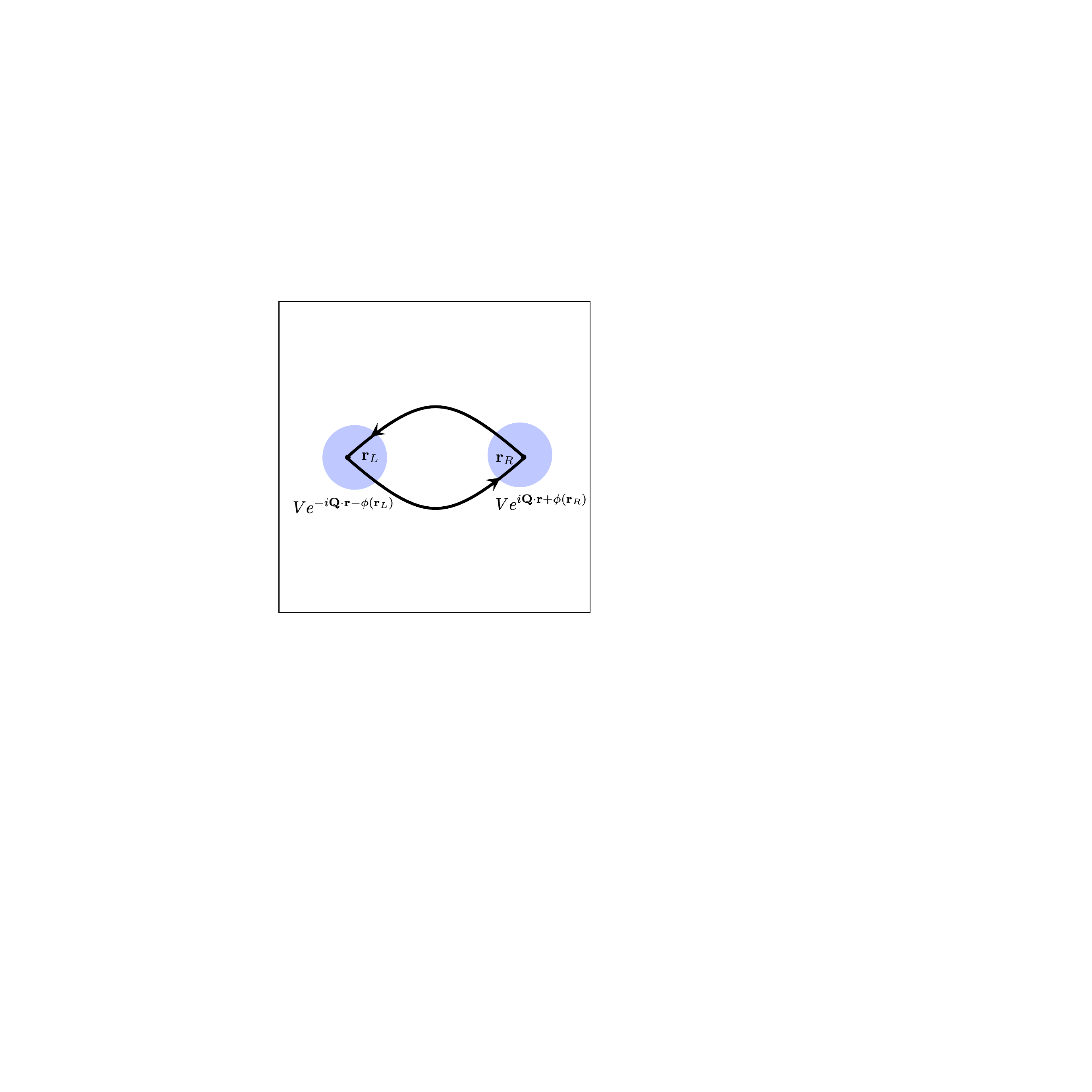}
\caption{\label{fig:scattering_potential} 
Effective scattering near real space scattering point.}
\end{figure}

We can understand this dependence as follows. According to scattering theory, the reflected wave is obtained by integrating the incoming wave against the scattering potential. Now let us consider the \emph{effective} scattering potential felt by a wave packet near $\bm{r}_{R/L}$, that is, within the blue regions (exaggerated in size for clarity) indicated in Fig.~\ref{fig:scattering_potential}. First consider $R$. The momentum transfer picks out the plus component of the CDW, and since the phase of the CDW is essentially constant in the blue region, we can make the replacement $\phi(\bm{r}) \to \phi(\bm{r}_R)$. Hence the effective potential is
\begin{equation}
U_R = Ve^{\bm{Q} \vdot \bm{r} + \phi(\bm{r}_R) }
\end{equation}
For $L$, the momentum transfer picks out the minus component, and we can make the replacement $\phi(\bm{r}) \to \phi(\bm{r}_L)$, giving
\begin{equation}
U_L = Ve^{-\bm{Q} \vdot \bm{r} - \phi(\bm{r}_L) }.
\end{equation}

It follows that
\begin{align}
\theta_R &= \phi(\bm{r}_R) + \ldots \\
\theta_L &= -\phi(\bm{r}_L) + \ldots \\
\end{align}
where the piece of the phase shift contained in $ \ldots$ is independent of the CDW phase. Putting these pieces together,
\begin{align}
S(E) &= S_{0}(E) + \phi(\bm{r}_R)  -\phi(\bm{r}_L) \\
&= S_{0}(E) + \Delta \phi
\end{align}
where $S_{0}(E)$ is the action for a uniform CDW. Since 
\begin{equation}
S_0(E) = \frac{\hbar A(E)}{eB} 
\end{equation}
with $A(E)$ the area of the stitched-together  $k$-space orbit (plus possible small corrections), we conclude that
\begin{equation}
S(E) = \frac{\hbar A(E)}{eB}  + \Delta \phi(\bm{R})
\end{equation}
where we have emphasized that the phase difference depends on the guiding center $\bm{R}$ of the orbit under consideration. This extra orbit-dependent phase will generically smear out the QO signal. Before discussing this point more quantitatively, we generalize this result to arbitrary dispersion and CDW order, and show that $\Delta \phi$ is a Berry phase.

\section{General result \label{sec:general_reconstruction}}
Consider a general dispersion and a CDW with ordering vectors $\bm{Q}_{1},  \ldots,\bm{Q}_M$, and corresponding phases $\phi_j(\bm{r})$. Generically, scattering across the Fermi surface of $H_0$ yields multiple closed orbits. Let us focus our attention on one of them. The different real-space scattering points can be labeled $\bm{r}_1,  \ldots \bm{r}_N$ (ordered sequentially). At scattering point $\alpha$, the wave packet scatters in $k$-space by some $\eta_{\alpha} \bm{Q}_{j_{\alpha}}$, where $\eta \in \{\pm 1\}$ and $j_{\alpha} \in \{1, \ldots M\}$. Then the effective scattering potential at scattering $\alpha$ is
\begin{equation}
U_{\alpha} = e^{i \eta_{\alpha} (\bm{Q}_{j_{\alpha}} \vdot \bm{r} + \phi_{j_{\alpha}}(\bm{r}_{\alpha}))},
\end{equation}
so we conclude 
\begin{equation}
S(E) = \frac{\hbar A(E)}{eB}  + \gamma(\bm{R}) \label{eq:general_result}
\end{equation}
with 
\begin{equation}
\gamma(\bm{R}) = \sum_{\alpha =1}^N \eta_{\alpha}\phi_{j_{\alpha}}(\bm{r}_{\alpha}).
\end{equation}

\section{Reconstructions in the cuprates \label{sec:cuprate_reconstruction}}

\begin{figure}
\includegraphics[width=6 cm]{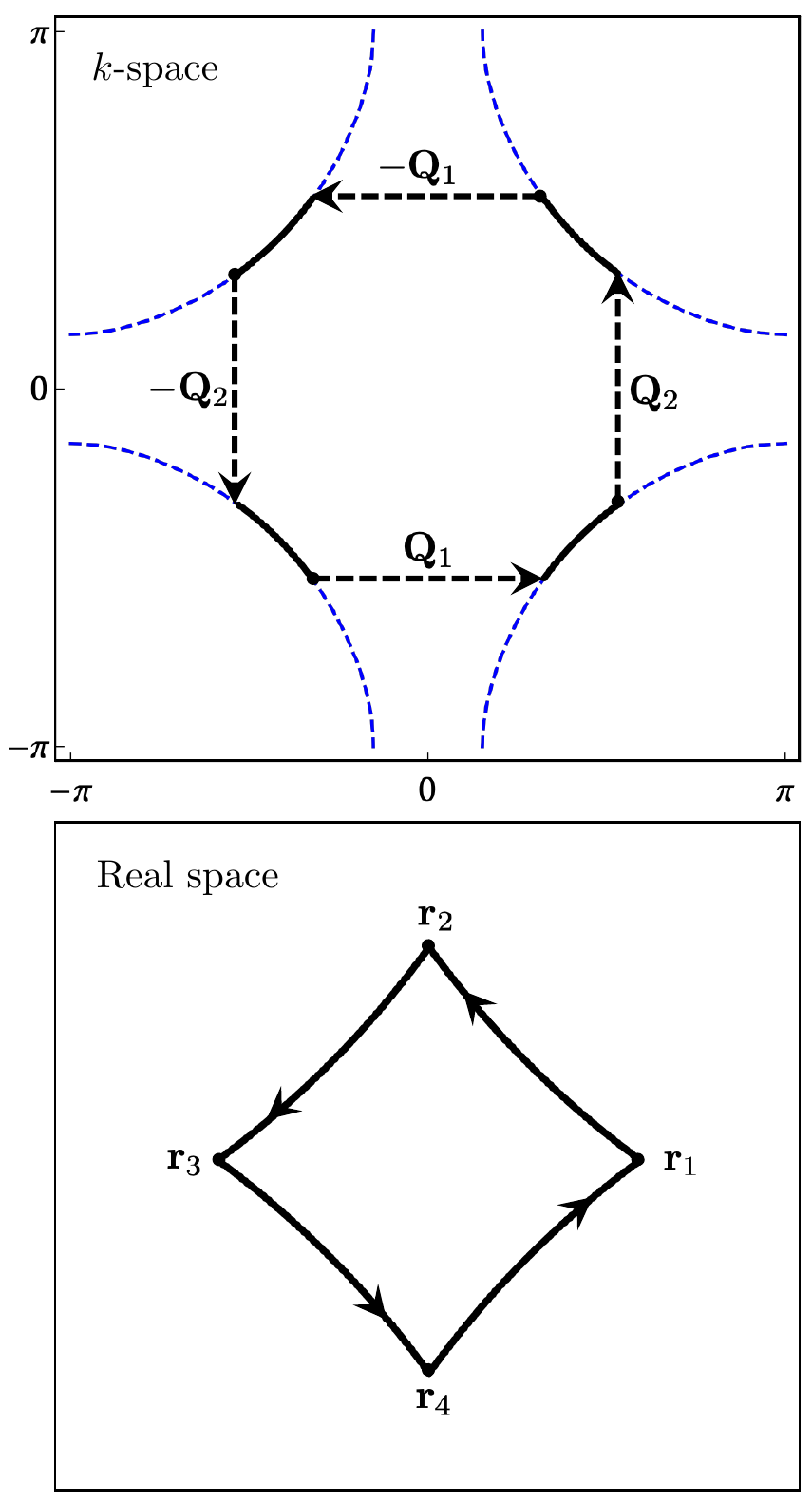}
\caption{\label{fig:bidirectional_reconstruction} 
Proposed reconstruction by bidirectional order in the cuprates.}
\end{figure}

Let us apply these results to the Fermi surface reconstructions proposed for the cuprates. Consider 
reconstruction by 
a bidirectional CDW \cite{Harrison_Sebastian_2011, Allais_et_al_2014} with $\bm{Q}_1 = (Q,0)$, $\bm{Q}_2 = (0,Q)$, and $Q \approx (1/3) (2\pi/a$), indicated in Fig~\ref{fig:bidirectional_reconstruction}.
Then in evaluating $\gamma$,
\begin{align}
j_1 &= j_3 = 1 \\
j_2&=j_4 = 2 \\
\eta_1 &= \eta_4 =  - \eta_2  = - \eta_3 = 1,
\end{align}
so
\begin{align}
\gamma(\bm{R}) &=   - [  \phi_1(\bm{r}_1) - \phi_1(\bm{r}_3) + \phi_2(\bm{r}_2) - \phi_2(\bm{r}_4)] \\
&= - [\Delta \phi_1 (\bm{R}) + \Delta \phi_2 (\bm{R})],
\end{align}
with $\Delta \phi_1 (\bm{R})$ the phase difference in component $1$ between right and left scattering points, and $\Delta \phi_2 (\bm{R})$ the phase difference in component $2$ between top and bottom.

\begin{figure}
\includegraphics[width=6 cm]{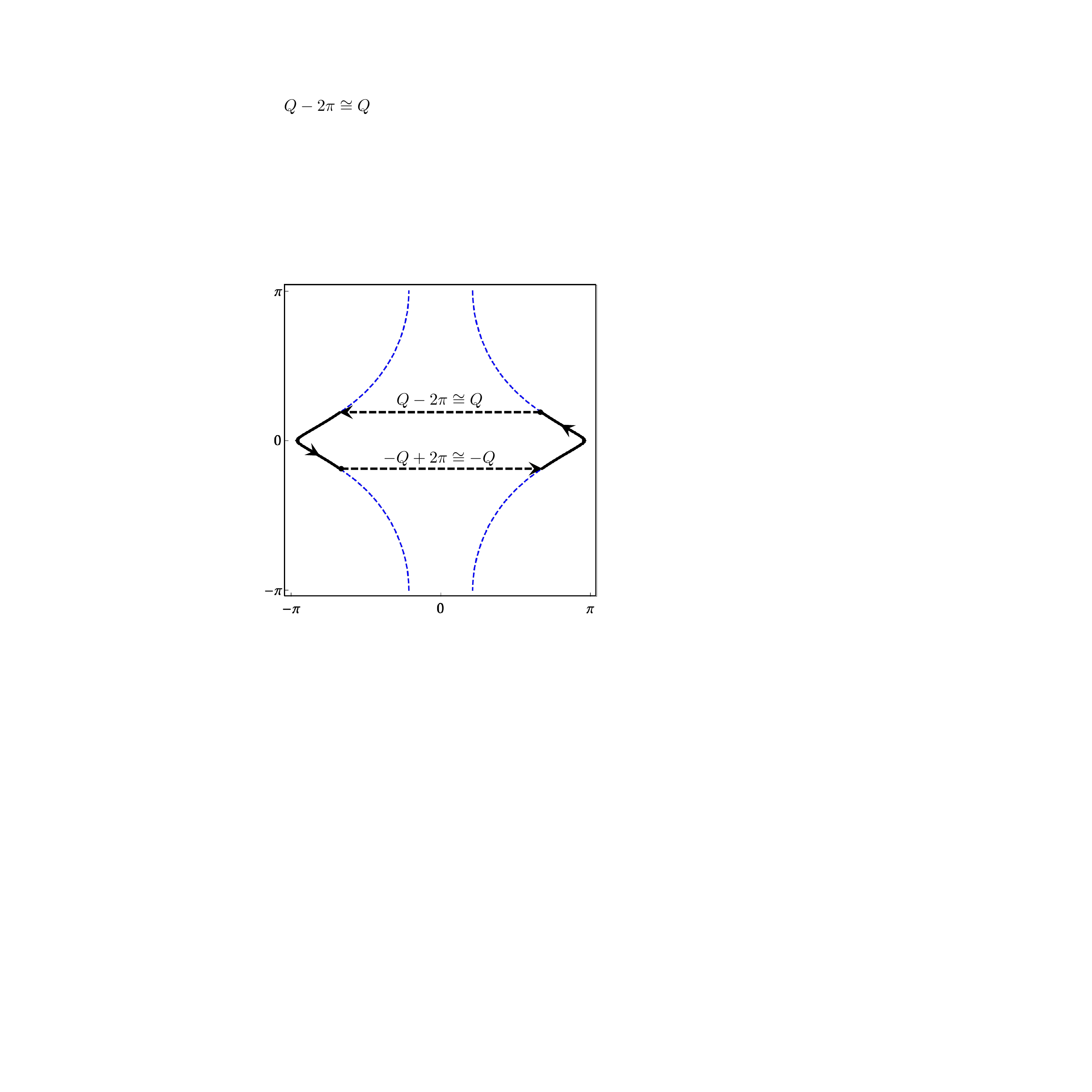}
\caption{\label{fig:unidirectional_reconstruction} 
Proposed reconstruction by unidirectional order in YBCO, with nematic distortion of the underlying Fermi surface.}
\end{figure}

Consider now reconstruction by a unidirectional CDW. 
This proposal presupposes a substantial nematic distortion of the underlying Fermi surface to 
obtain the 
requisite electron-like pocket \cite{Yao_Lee_Kivelson_2011}. The resulting reconstruction is indicated in Fig.~\ref{fig:unidirectional_reconstruction}. Since the topology of the reconstructed orbit is the same as in the model Hamiltonian, $\gamma(\bm{R}) = \Delta \phi (\bm{R})$ as before.

\section{Berry phase approach \label{sec:berry}}
Here we solve the problem using a fully semiclassical approach, accounting for the CDW as part of the band structure. In this picture, the extra phase $\gamma$ appears as a Berry phase. 

We follow the general theory of Berry phases in solid-state systems developed by Sundaram and Niu\cite{Sundaram_Niu_1999}. For the sake of definiteness, consider the model Hamiltonian. Drawing on insight from the scattering picture, we construct a wave packet from the Bloch-like states
\begin{equation}
\ket{\psi_{\bm{k}}(\bm{r})} = c_{1}(\bm{k},\bm{r}) \ket{\bm{k}} + c_{2}(\bm{k},\bm{r}) \ket{\bm{k}+\bm{Q}} \label{eq:bloch_like_state},
\end{equation}
which diagonalize the projection of the local Hamiltonian near $\bm{r}$ onto the span of $\{\ket{\bm{k}}, \ket{\bm{k} + \bm{Q}  }\}$. Explicitly, the vector of coefficients $c = (c_1,c_2)$ satisfies $\mathcal{H}c= \mathcal{E}c$, where the effective Hamiltonian is
\begin{equation}
\mathcal{H}(\bm{r},\bm{k}) = 
\begin{pmatrix} 
\mathcal{E}_0(\bm{k}) & Ve^{-i\phi(\bm{r})} \\
Ve^{i\phi(\bm{r})} & \mathcal{E}_0(\bm{k}+\bm{Q}),
\end{pmatrix} \label{eq:H_eff}
\end{equation}
and $\mathcal{E}$ is the local band energy. In the uniform limit, the upper band yields the reconstructed Fermi surface. 

The band energy $\mathcal{E}$ is unaffected by CDW phase disorder, so the leading order effect is entirely due to Berry phase terms. These appear in the classical Lagrangian which governs the evolution of $\bm{r}$ and $\bm{k}$. Its calculation proceeds exactly as in Ref.~\onlinecite{Sundaram_Niu_1999}, except the cell periodic function is everywhere replaced by $c$. This Lagrangian, along with further elaboration on the Berry phase approach, are presented in Appendix~\ref{app:berry_curvature}.

The Berry phase terms imply that the total action is now a sum of $\hbar A(E_F)/(eB)$ and a Berry phase $\gamma$, which we compute here directly from the Berry connection. In Appendix~\ref{app:berry_curvature}, $\gamma$ is obtained from the associated Berry curvature.

Let us parameterize the orbit by ${t \in [0,T]}$, where $T$ is the cyclotron period, and consider the total Berry connection
\begin{equation}
\mathcal{A}(t) = \mel**{c(\bm{r},\bm{k})}{i \dv{t}}{c(\bm{r},\bm{k})},
\end{equation}
where bra-ket notation involving $c$ means the usual inner product on $\mathbb{C}^2$. We are free to demand that $c \approx (1,0)$ on the right half of the Fermi surface and $c \approx (0,1)$ on the left half of the Fermi surface. In this partially fixed gauge $\mathcal{A}(t)$ vanishes except at the left and right scattering points. Then the Berry phase is
\begin{equation}
\gamma = \int_0^T dt \hspace{ 1 pt} \mathcal{A}(t) = 
\int_{I_L} dt \hspace{ 1 pt} \mathcal{A}(t) 
+  \int_{I_R} dt \hspace{ 1 pt}  \mathcal{A}(t)
\end{equation}
where $I_{L/R}$ are small time intervals about the left and right scattering points.

Consider the integral over ${I_R = [t_i,t_f ]}$.  The boundary conditions are $c(t_i) \approx (1,0)$, $c(t_f) \approx (0,1)$. In addition, the phase is essentially constant over this interval, so we can substitute $\phi(\bm{r}) \to \phi(\bm{r}_R)$ in Eq.~(\ref{eq:H_eff}). As shown in Appendix~\ref{app:berry_avoided_crossing}, this fixes the value of the integral:
\begin{equation}
\int_{I_R} dt  \mathcal{A}(t) = \phi(\bm{r}_R).
\end{equation}
Similarly,
\begin{equation}
\int_{I_L} dt  \mathcal{A}(t) = -\phi(\bm{r}_L)
\end{equation}
so that $\gamma = \Delta \phi$, reproducing Eq.~(\ref{eq:general_result}). More generally, either the CDW or the underlying band structure may break inversion symmetry. Then a Berry phase appears even in the uniform limit, and $\gamma$ as defined in Eq.~(\ref{eq:general_result}) is the piece induced by CDW disorder. Above, we neglected the modification to the semiclassical trajectories; this is justified in Appendix~\ref{app:action_perturbation}.

In a recent paper on semiclassical dynamics in quasicrystals \cite{Spurrier_Cooper_2018}, Spurrier and Cooper have also obtained an expression for the Berry phase as a sum of Bragg scattering phase shifts. In that context, a non-zero Berry phase appears because the Bragg scattering points may be connected non-trivially in momentum space -- whereas here, it is a consequence of a spatially varying CDW phase.

\section{Dingle Factors \label{sec:dingle}}
In this section, we find the Dingle factor and Dingle field. In a uniform system, the density of states is
\begin{equation}
\nu(E_F,B) = \frac{A'(E_F)}{(2\pi)^2} \sum_p (-1)^p \exp
\left[ip S(E_F)\right]
\label{eq:DOS_FT}.
\end{equation}
In a disordered system, we should average this over all orbits, so that
\begin{multline}
\nu(E_F,B) \\
 =\frac{A'(E_F)}{(2\pi)^2} \sum_p (-1)^p R_D(p) \exp[ip \pqty{\frac{2\pi F}{B} + \theta(p) } ]. \label{eq:DOS_disorder_FT}
\end{multline}
where 
\begin{align}
R_D(p) &= \abs{\overline{e^{ ip \gamma}}} \\
\theta(p) &= \arg(\overline{e^{ ip \gamma}})
\end{align}
and the bar denotes an average over different orbits.

\subsection{Fundamental}
Here we consider the fundamental, $R_D \equiv R_D(p\!=\!1)$. 
In the unidirectional case,
\begin{equation}
    R_{D,1Q} = \abs{\overline{e^{i\Delta \phi }}} 
\end{equation}
is the CDW correlation function evaluated at the distance between left and right scattering points. Letting $2R_c$ denote this distance,
\begin{equation}
    R_{D,1Q} \sim e^{-2R_c/\xi}.
\end{equation}
In the bidirectional case
\begin{equation}
    R_{D,2Q} = \abs{\overline{e^{i(\Delta \phi_1 + \Delta \phi_2)}}}.
\end{equation}
Assuming the two CDW components are independent, this factors into a product of two correlation functions, so  
\begin{align}
    R_{D,2Q} &\sim e^{-2R_c/\xi} \times e^{-2R_c/\xi} \\ 
    &\sim e^{-4R_c/\xi}
\end{align}
where $2R_c$ is again the distance between left and right scattering points, or, equivalently, top and bottom scattering points. 

In terms of the Dingle field, where $R_D = e^{-B_D/B}$, 
\begin{equation}
    B_D = \frac{ 2n \hbar k_F}{e\xi}, \label{eq:B_D}
\end{equation}
where $n = 1$ for the unidirectional reconstruction, $n=2$ for the bidirectional reconstruction, and $2k_F$ is the distance between scattering points in momentum space.

\subsection{Higher harmonics}
\label{sec:higher harmonics}
The result for the first harmonic depends only on the correlation length, and not on the microscopic details of the CDW -- in particular, it does not matter whether the phase fluctuates smoothly throughout the sample, or if there is instead a random array of sharp DCs.

The higher harmonics are, however, sensitive to this distinction. In the smoothly fluctuating case, $R_D(p)$ decreases rapidly with $p$. For instance, if $\Delta \phi$ is Gaussian distributed, then 
\begin{equation}
R_D(p) = \exp(-2p^2R_c/\xi).
\end{equation}
Note that in contrast to what is expected for potential scattering \cite{shoenberg2009magnetic} the exponent in the Dingle factor is quadratic in $p$, not linear.

For a random DC array, however, certain harmonics are completely unaffected by phase disorder. Consider, for the sake of definiteness, a unidirectional reconstruction (in either our model Hamiltonian or YBCO) with local wave vector $Q=  (p/q)(2\pi/a)$, with $p$, $q$ relatively prime positive integers. Then the phase difference between any two points in the sample is  $\Delta \phi = 2\pi n/q$ for some integer $n$. Therefore $R_D(p^{\star})=1$ whenever $p^{\star}$
is an integer multiple of $q$. All other harmonics are generically damped. As a consequence, higher harmonics can be significantly stronger than the fundamental. 

One way understand this result is that after $q$ periods all electrons in the sample have gained the same phase modulo $2\pi$. The signal may also be understood as a superposition of conventional QO signals offset by integer multiples of $1/q$ times the period, leading to multiple peaks per oscillation period instead of de-phasing. Below, we examine the real-space Landau level spectrum in the presence of DCs, showing explicitly how different parts of the sample contribute to the density of states.

\section{Numerics \label{sec:numerics}}

In this section, 
we test (and confirm) the above predictions by exact numeric experiments on our model Hamiltonian. 
\subsection{Numeric technique and setting up the model}
We use a 
recursive Green's function method, which allows computing the density of states of
1-D systems with computational effort scaling linearly in length \cite{Lee_Fisher_1981,Yi_Maharaj_Kivelson_2015, Allais_et_al_2014}. By choosing the Landau gauge $\bm{A} = (0,Bx)$, and considering a CDW that is perfectly correlated in the $y$-direction, we can preserve translation symmetry in the $y$-direction. Therefore we can reduce the problem to independent 1-D chains labeled by canonical momentum $p_y$.

Explicitly, we start with
\begin{multline}
H = \sum_{\bm{r}} \Big[
- \pqty{ c^{\dagger}_{\bm{r} + \bm{\hat{x}}} c_{\bm{r}}
+  e^{-i 2\pi B x} c^{\dagger}_{\bm{r} +\bm{\hat{y}}}c_{\bm{r}} + \text{H.c}} \\
+ 2V\cos[Qx + \phi(x)]c^{\dagger}_{\bm{r}}c_{\bm{r}} \Big]
\end{multline}
For this section, we use units where $a=t=h/e=1$.

Introducing the operator $c^{\dagger}_{x,p_y}$ which creates a state localized in the $x$-direction and with crystal momentum $p_y$ in the $y$-direction,
\begin{multline}
H = \sum_{p_y} \sum_{x} \Big[
 - \pqty{ c^{\dagger}_{x+1,p_y} c_{x,p_y} + \text{H.c}} \\
+ \big( -2\cos[p_y+ 2\pi B x]  \\
+2V\cos[Qx +\phi(x)] \big) c^{\dagger}_{x,p_y} c_{x,p_y}  \Big]
\label{eq:numerics_input}
\end{multline}

In Ref.~\onlinecite{Yi_Maharaj_Kivelson_2015}, it is shown that for a uniform CDW, for generic $B$ the density of states is independent of $p_y$ in the thermodynamic limit. This is because the role of $p_y$ is just to specify the minima of the effective cosine potential in the second line of Eq.~(\ref{eq:numerics_input}), but the offset of the minima relative to the lattice will drift throughout the sample even if we do not average of $p_y$. For a disordered CDW, the density of states is still independent of $p_y$, by self-averaging. 

We generate the phase $\phi(x)$ of the CDW by computing a 1-D random walk. To generate  a smoothly varying phase, we allow the increment at each step to be Gaussian distributed. For a random DC array, we allow only a discrete jump at each step. In each case the resulting phase is smoothed out.

\subsection{Dingle field for smoothly varying phase}
Here we present results for a smoothly varying phase;  we extract values of $B_D$, and compare them with the theoretical predictions.

We set $Q = 2\pi/3$ and consider $2V = 0.175$ and $2V = 0.11$, values that compromise between maintaining well-defined Bragg scattering points and avoiding magnetic breakdown.
We compute the density of states at a fixed energy, $E = -0.22 $, as a function of $B$.
To confirm that the density of states is sufficiently insensitive to the absolute phase of the CDW even in this commensurate case, we add a small linearly varying phase and find a slight shift in the QO frequency, but no damping.

\begin{figure}
\includegraphics[width=8.75 cm]{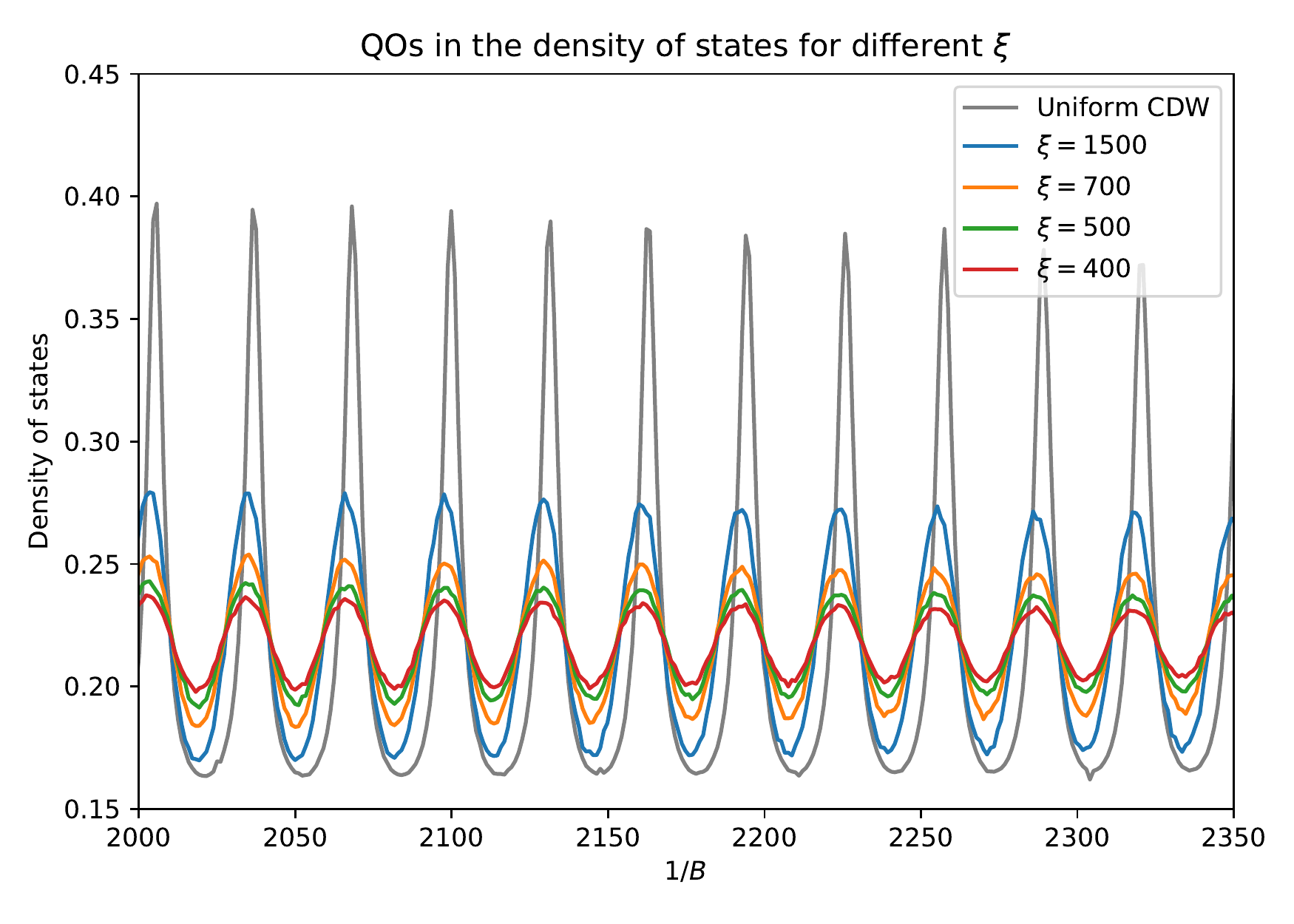}
\caption{\label{fig:dos_various_xi} 
Numerically computed QOs in the density of states for several values of $\xi$. Parameter values indicated in the main text. }
\end{figure}

For $2V = 0.175$, results for CDW correlation length $\xi = 1500,700,500$ and $400$ are shown in Fig.~\ref{fig:dos_various_xi}, together with results for a uniform CDW.
Note that we added a small imaginary part to the energy in order to broaden out singularities. As a result, the signal amplitude decreases with $1/B$ even for the uniform CDW.

To extract the numerically observed Dingle field $B_D^{\text{num}}$, we Fourier transform over several $1/B$ windows, obtaining the $1/B$ dependence of the amplitude of the fundamental. The amplitude at each field is normalized by the amplitude at the same field for a uniform CDW, and the log of the result is fit to a straight line. Comparison with the  predicted Dingle field (Eq.~\ref{eq:B_D}) is shown in Table~\ref{tab:compare_th_num}. The results are in very good agreement; the consistent underestimate for $2V = 0.175$ reflects the relatively large CDW amplitude, which leads to additional de-phasing effects not captured by our theory.
\begin{table}[h]
\centering
\begin{tabular}{|c|c|c|c|} 
    \hline
$\xi$ & $B_D \times 10^{4} $  & $B_D^{\text{num}} \times 10^{4}$ & $B_D^{\text{num}} \times 10^{4}$   \\
&  &   ($2V = 0.175$) & ($2V = 0.11$) \\
\hline
$1500$ & $1.94 $ & $2.12 \pm 0.02 $ & $1.93 \pm 0.02$  \\
$700$ & $4.16 $ & $4.47 \pm 0.04 $ & $4.14 \pm 0.05$ \\
$500$ &$5.82 $ & $6.22 \pm 0.03 $ & $5.81 \pm 0.05$ \\
$400$ &$7.28 $ & $7.78 \pm 0.04 $ &  $7.33 \pm 0.06$ \\
\hline
\end{tabular}
\caption{Comparison of theoretically expected Dingle field $B_D$, and numerically observed Dingle field $B_D^{\text{num}}$. Error  bars are the standard error of the least-squares fit. \label{tab:compare_th_num}}
\end{table}

\subsection{Persistent oscillations for a sharp array of DCs}
\begin{figure}
\includegraphics[width=8.75 cm]{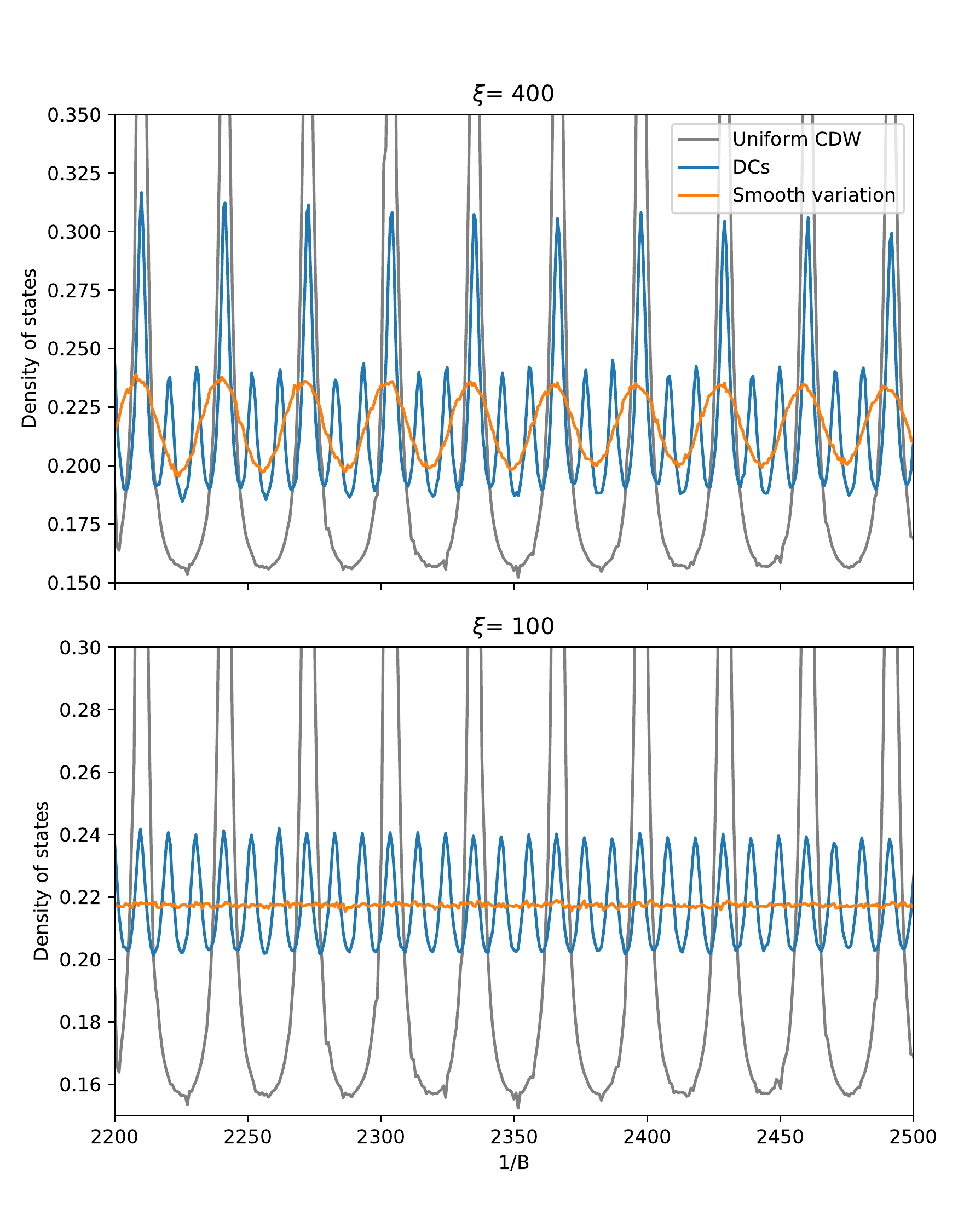}
\caption{\label{fig:comparison} 
Comparison of QOs for disordering via DCs and smoothly varying phase, for $\xi = 400$ (top) and $\xi = 100$ (bottom)}
\end{figure}

Here we present results for a random array of sharp DCs. Still with $Q = 2\pi/3$, we generate a correlation length $\xi$ by an appropriate density of $\pm Q$ phase DCs.
QOs for $\xi = 400$, $100$ are shown in Fig.~\ref{fig:comparison}, where they are compared against a CDW with the same correlation length but a smoothly varying phase. For $\xi = 400$, the harmonic content for DCs is already highly unusual; by $\xi = 100$ the oscillations have apparently tripled in frequency while oscillations in the smoothly varying case are undetectable. This is precisely what was predicted in Sec.~\ref{sec:higher harmonics}.

\subsection{Real-space Landau  level structure near a discommensuration}
Here we demonstrate the real-space Landau level structure near a DC. Consider a commensurate CDW with $Q = 2 \pi (p/q)$. Take a phase configuration $\phi(x)$ such that $\phi(x) = 0$ for $x<0$, and $\phi(x) = \pm 2\pi/q$ for $x>0$ -- that is, a single, (minimal) DC at $x=0$. Then for all Landau levels with guiding center $R_x$ such that $\abs{R_x} > R_c$ $\gamma = 0$, but for the remaining Landau levels with $\abs{R_x} < R_c$, $\gamma = \pm2 \pi / q$. Aside from the difference in $\gamma$, nearly every orbit is unaffected by the DC. Individual guiding-center resolved energy levels are well defined, with quantization condition
\begin{equation}
\frac{\hbar A(E_n)}{eB}  + \gamma(\bm{R}) = 2 \pi \pqty{n + \frac{1}{2}}, \label{local_quantization}
\end{equation}
so Landau levels with $\abs{R_x} > R_c$ are unaffected by the DC, but all Landau levels with $\abs{R_x} < R_c$ are shifted in energy by $\mp \hbar \omega_c /q$.

\begin{figure}
\includegraphics[width=9.75 cm]{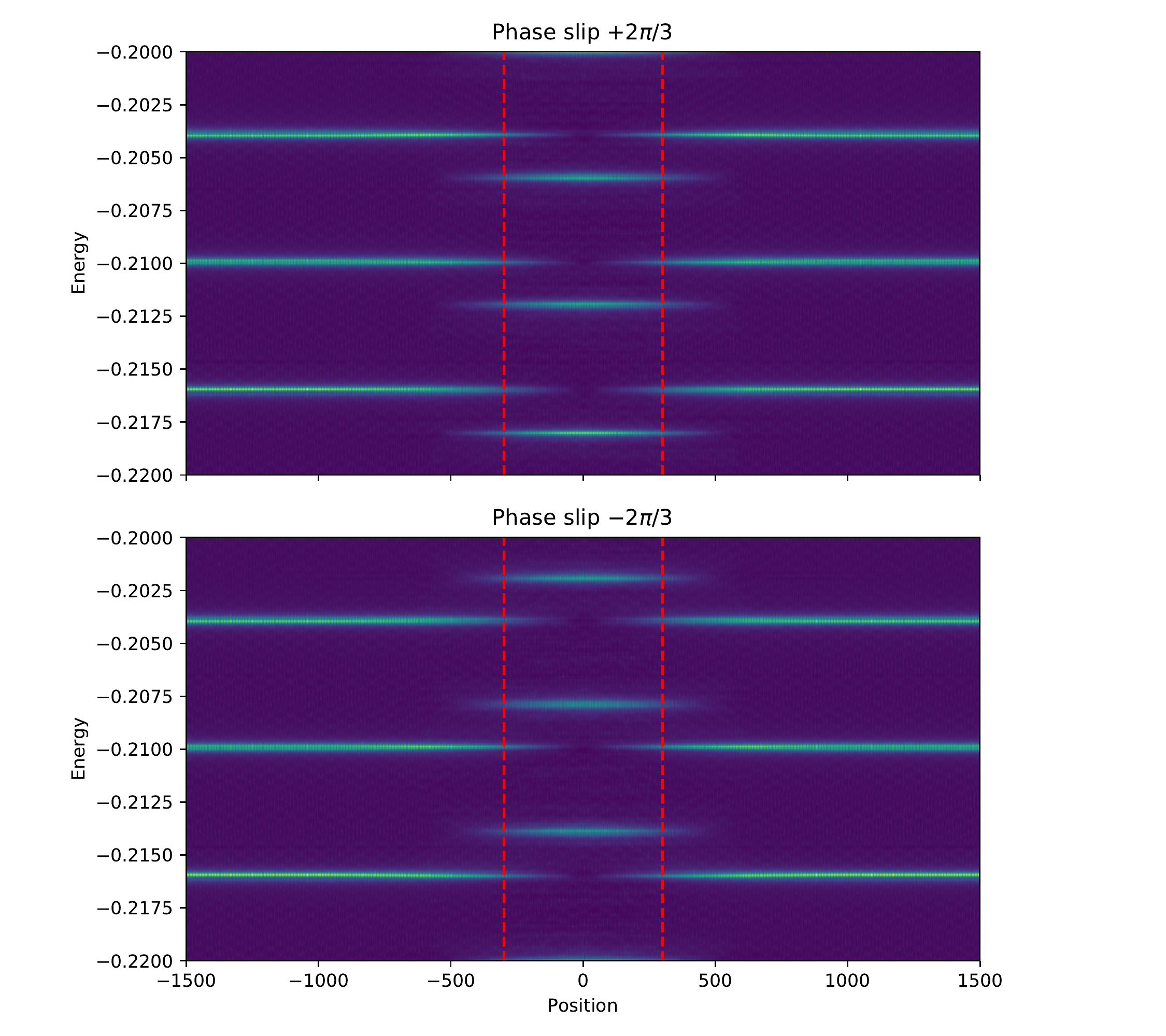}
\caption{\label{fig:real_space} 
Position resolved density of states with a single DC at $x=0$, whose phase slips by $+2\pi/3$ (top) or $-2\pi/3$ (bottom) as $x$ increases. Red dashed lines mark $\pm R_c$.}
\end{figure}

Taking $Q= 2 \pi/3$, we demonstrate this by computing the position-resolved density of states for a single DC in Fig.~\ref{fig:real_space}. The levels are shifted through $\pm 1/3$ of their spacing, with the expected sign. Note that for a sharp DC, the spectrum in the vicinity of the DC does not bend down to meet the spectrum away from the DC; instead, density of states from the shifted levels dies off 
while density of states from the un-shifted levels simultaneously picks up. This is because a Landau level with guiding center $R_x$ contributes density of states for all $x \in [ R_x-R_c,R_x+R_c ] $.

This striking Landau level structure is a consequence of the  Berry phase being uniquely defined by the phase difference between the two scattering points; the effect should be observable in scanning tunneling spectroscopy on a suitable CDW system, similar to recent experiments which observe defect-shifted Landau levels \cite{Feldman_et_al_2016}.

\section{Application to YBCO \label{sec:experiments}}

Here we apply the present theoretical results to the case of the QOs seen in \ch{YBa2Cu3O_{$y$}}.  It is a rational supposition, but not one which is directly confirmed in experiment, that the small Fermi surface areas apparent in the QOs reflect a Fermi surface reconstructed by one or the other of the observed CDW orders (phenomenological descriptions of these orders may be found in Ref.~\onlinecite{Jang_et_al_2016, Caplan_et_al_2015, Caplan_Orgad_2017}). However, given that the CDW correlation lengths are at best comparable to the cyclotron radius, it is worth asking whether the observed Dingle factors are consistent with these correlation lengths. In this context, we consider the Dingle factors computed above to represent an upper bound on the QO amplitude -- other forms of disorder (in addition to that represented by the phase-disordering of the CDW) will only reduce the amplitude of the QOs further.

We denote by $B^{\text{exp}}_{\text{D}}$ the actual Dingle field observed in QO experiments on \ch{YBa2Cu3O_{$y$} }.
Consistency requires 
\begin{equation}
B^{\text{exp}}_{D} > B_{D}, \label{eq:inequality}
\end{equation}
where CDW phase disorder alone would produce the Dingle field
\begin{equation}
    B_D = \frac{ 2n \hbar k_F}{e\xi}, 
\end{equation}
where for unidirectional (bidirectional) order ${n= 1}$ (${n=2}$) and $\xi = \xi_{1Q}$ ($\xi = \xi_{2Q}$) is the relevant correlation length extracted from x-ray data. Note that since $n=1$ in the unidirectional case whereas $n=2$ in the bidirectional case, we expect more strongly damped QOs in the bidirectional case, even before taking into account the difference in correlation length.
\subsection{Evaluating $B_D$}
First, we rewrite the above in terms of the QO frequency $F$ as
\begin{equation}
    B_D = \frac{2n \alpha}{\xi} \sqrt{\frac{2\hbar F}{e}}
\end{equation}
where $\alpha$ depends on the geometry of the reconstructed orbit, and is defined by the relation 
\begin{equation}
    k_F = \alpha \sqrt{A(E)/\pi},
\end{equation}
so that for a circular Fermi surface, $\alpha = 1$. For the diamond-like orbits in Fig.~\ref{fig:bidirectional_reconstruction} and \ref{fig:unidirectional_reconstruction}, the value $\alpha \approx 1.25 $ would be appropriate. However, since the precise shape of the reconstructed Fermi surface is at present unknown, we opt for a face-value analysis with $\alpha = 1$. 

For bidirectional order, Ref.~\onlinecite{Chang_et_al_2012} finds ${\xi_{2\text{Q}} \approx 100 \text { \AA}}$ in $y = 6.67$, at $17 \text{ T}$ and $2 \text{ K}$. For unidirectional order, Ref.~\onlinecite{Chang_et_al_2016} finds ${\xi_{1\text{Q}} \approx 190 \text { \AA}}$ in ${y = 6.60}$ and ${\xi_{1\text{Q}} \approx 310 \text { \AA}}$ in ${y = 6.67}$, both at $\approx 17 \text{ T}$.
With these correlation lengths, $F \approx 540 \text{T}$, and $\alpha=1$, we find
\begin{align}
B_{D,2Q} & \approx
\begin{aligned}
340 \text{ T} && y = 6.67 
\end{aligned} \\
B_{D,1Q} &\approx 
\begin{cases}
90 \text{ T} & y = 6.60 \\
55 \text{ T} & y = 6.67 
\end{cases} \label{eq:B_D_lower_bound}
\end{align}
Uncertainty in these values is due to uncertainty in $\alpha$ and ambiguity in the way that $\xi$ is extracted from the structure factor; the references above define $\xi$ as the inverse standard deviation of a Gaussian fit.

\subsection{Measured Dingle field}
In Table~\ref{tab:B_D_values}, we report experimental Dingle field measurements in \ch{YBa2Cu3O_{$y$}} for several dopings. 
\begin{table}[h]
\centering
\begin{tabular}{|c|c|c|} 
    \hline
$y$ & $p$ (doping) & $B^{\text{exp}}_D$ (T)  \\
\hline
$6.51$ & 0.092 & $260 \pm 25$ \\
$6.59$ & 0.110 & $110 \pm 20$ \\
$6.67$ & 0.125 & $420 \pm 40$ \\
$6.75$ & 0.135 & $460 \pm 45$ \\
$6.80$ & 0.140 & $560 \pm 55$ \\
$6.86$ & 0.152 & $750 \pm 75$ \\
\hline
\end{tabular}
\caption{Observed Dingle fields $B^{\text{exp}}_D$. The error bars are represent the uncertainty due to multiple oscillation frequencies. Data from Ref. \onlinecite{ramshaw2015quasiparticle,ramshaw2012shubnikov}. \label{tab:B_D_values}}
\end{table}

\begin{figure}
\includegraphics[width=9.0 cm]{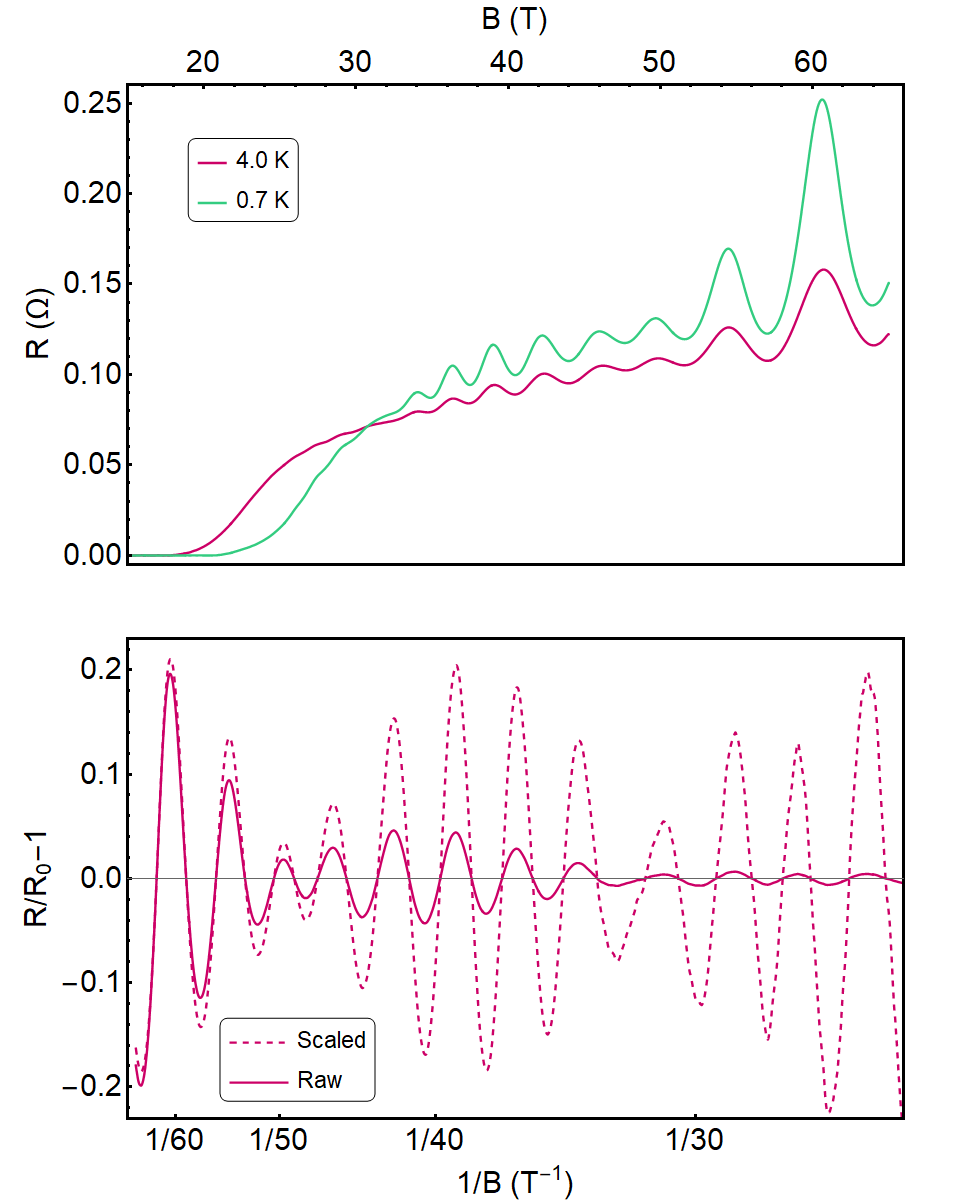}
\caption{\label{fig:data} 
Raw (top) and background-removed (bottom) data for $y=6.59$. We extract $B_D$ from the 4 K data, where the second harmonic is suppressed by temperature. We divide out the factor $R_T = \frac{2 \pi^2 k_B T}{\hbar \omega_c}/\sinh(\frac{2 \pi^2 k_B T}{\hbar \omega_c})$, where the cyclotron frequency $\omega_c = eB/m^{\star}$ is known from previous measurements \cite{Ramshaw_et_al_2011}. The data is then scaled by $e^{+B_D/B}$, with $B_D$ chosen so that the amplitude is field-independent.}
\end{figure}

Except for $y=6.59$, the Dingle field is extracted by fitting the background-removed data to a Lifshitz-Kosevich (LK) form. For $y=6.59$ -- where the strongest oscillations are observed down to the lowest fields -- the presence of multiple frequencies of comparable amplitude complicates the fit. Instead, we first divide out the known temperature-dependent LK factor, and then scale the oscillations by a factor $e^{+B_D/B}$, choosing $B_D$ so that the amplitude of the signal is as constant as possible over the relevant field range. This procedure is presented, along with the raw data, in Fig.~(\ref{fig:data}).

\subsection{Consistency check}
Consider first reconstruction via bidirectional order. Zero field measurements indicate $\xi_{2Q}$ does not depend strongly on doping \cite{Blanco-Canosa_et_al_2014, Hucker_et_al_2014}; assuming this holds in-field as well, we compare all QO measurements against ${B_{D,2Q} \approx 340 \text{ T}}$. While the lower bound in Eq.~(\ref{eq:inequality}) is satisfied by most dopings, it is violated by $B_D^{\text{exp}} = 110 \text{ T}$ in $y = 6.59$. This violation is quite severe, as the Dingle field enters the amplitude through an exponent: the observed QO signal drops in amplitude by a factor of $50$ going from $70 \text{ T}$ down to $20 \text{ T}$, whereas the predicted drop assuming a bidirectional reconstruction is by a factor of at least $2 \times 10^{5}$. We conclude that it is not easy to reconcile the experimental observations with reconstruction by the bidirectional CDW.

Consider now reconstruction by unidirectional order, where we compare $B_D^{\text{exp}}$ in $y = 6.67$ against $B_{D,2Q}$ in the same doping, and $B_D^{\text{exp}}$ in $y = 6.59$ against $B_{D,2Q}$ in $y = 6.60$. The lower bound in Eq.~(\ref{eq:inequality}) is satisfied in both cases, although it comes close saturation in $y = 6.59$. Reconstruction by the unidirectional CDW is therefore at least marginally consistent; the near saturation in $y = 6.59$ could indicate this doping has very little disorder besides CDW phase disorder.

The above analysis relied on correlation lengths measured at only $17 \text{ T}$, but pulsed-field measurements (which are less accurate, but can access higher fields) indicate $\xi_{2Q}$ is roughly constant above $15 \text{ T}$, whereas $\xi_{1Q}$ grows \cite{Gerber_et_al_2015, Jang_et_al_2016}. Therefore, considering the field dependence of $\xi$ does not change the conclusions above.

\subsection{Interpreting the main peak as a harmonic}
Soon after quantum oscillations were discovered in YBCO it was pointed out that three primary frequencies are apparent in the Fourier transform of the oscillations \cite{Audouard_et_al_2009}. The main frequency is approximately $540 \text{ T}$, with two side lobes situated at approximately $\pm 90 \text{ T} $. Measurements over a broader field range reveal a second set of side lobes, situated a further $\pm 90 \text{ T}$ from the first two side lobes \cite{Maharaj_et_al_2016}. Combined with the reported solitary frequency at $95 \pm 10 \text{ T}$ \cite{Doiron_et_al_2015}, this presents the unusual scenario of 6 oscillation frequencies that are all multiples of approximately $90 \text{ T}$.

Our result that certain harmonics of the quantum oscillations can be undamped by phase disorder raises the interesting possibility that \textit{all} observed frequencies are harmonics of a single pocket with area $\approx 90 \text{ T}$. As shown in section \ref{sec:higher harmonics}, a random array of DCs in a locally-period-3 CDW would leave the 3rd and 6th harmonics undamped. Indeed, the 6th harmonic-- $540 \text{ T}$ --is the dominant frequency observed. The lack of a significant 3rd harmonic could be explained by considering the Zeeman-splitting factor of QOs, or the inclusion of other damping coefficients including those due to regular disorder and temperature. It is therefore not inconceivable that the complex but regularly-spaced spectral structure in the QO Fourier transform, and the dominance of the $540 \text{ T}$ frequency, is due to the special damping factor from random DCs in a locally-period-3 CDW. Indeed, random DCs have been directly observed in the cuprate BSCCO using scanning tunneling microscopy\cite{Mesaros_et_al_2016}, making it plausible that the CDW in YBCO has the requisite microscopic character.

One appealing corollary of a small fundamental pocket size in YBa$_2$Cu$_3$O$_{6.59}$ pertains to the heat capacity. The 95 tesla pocket reported by \citet{Doiron_et_al_2015} has a cyclotron mass of only $m^{\star} = 0.45 \pm 0.1 ~m_e$---not too far from 1/6th of the mass measured in YBa$_2$Cu$_3$O$_{6.59}$ ($m^{\star}/6 \approx 0.3~m_e$ \cite{Ramshaw_et_al_2011}). 
Assuming a single pocket per copper oxide plane, with two planes per unit cell in YBCO, the electronic contribution to the heat capacity would be $\gamma = 0.9$~mJ/mol K$^2$.
As open sheets would further contribute to the heat capacity but would not be seen in QOs, this is consistent with the experimental value of $\approx 4$~mJ/mol K$^2$ observed in the normal state of YBa$_2$Cu$_3$O$_{6.59}$ in high magnetic fields\cite{Riggs_et_al_2011,Marcenat_et_al_2015}.

\begin{acknowledgments}
We would like to acknowledge helpful discussions with Johan Chang, Jing-Yuan Chen, Benjamin E. Feldman, Xiao-Qi Sun, and Louis Taillefer. This work was supported, in part, by NSF grant \# DMR-1608055 (SAK and YG).
\end{acknowledgments}

\appendix
\section{Magnetic breakdown \label{app:mb}}
Here we derive Eq.~(\ref{eq:field_regime}). At an $n$th order scattering point, the CDW opens a gap $\Delta$ between $\mathcal{E}_0(\bm{k})$ and ${\mathcal{E}_0(\bm{k} \pm n\bm{Q})}$. Magnetic breakdown near the avoided crossing may be neglected as long as
 \begin{equation}
    \frac{eB}{\hbar} \ll \pqty{\frac{\Delta}{v_F}}^2
\end{equation}
is satisfied\cite{Ashcroft_Mermin_1976}. Conversely, there is nearly full MB if the reverse inequality is satisfied. Moreover, ${\Delta \sim V (V/E_F)^{n-1}}$ at an $n$th order scattering point, so it is possible to have negligible magnetic breakdown at $n$th and lower order scattering points but nearly full magnetic breakdown at $(n+1)$th and higher order scattering points if
\begin{equation}
    \pqty{\frac{V}{v_F}}^2 \pqty{ \frac{V}{t} }^{2n}
    \ll \frac{eB}{\hbar} 
    \ll \pqty{\frac{V}{v_F}}^2 \pqty{ \frac{V}{t} }^{2n-2}.
\end{equation}
Similar conditions were invoked in Ref.~\onlinecite{Yi_Maharaj_Kivelson_2015, Spurrier_Cooper_2018}.
Choosing the case $n = 1$ above yields Eq.~(\ref{eq:field_regime}). 

\section{Details of the semiclassical formalism \label{app:berry_curvature}}
Here we discuss the details of the semiclassical formalism as applied to our model Hamiltonian. We work in units where $\hbar = 1$.

\subsection{General formalism}
The Lagrangian which generates the EOM Eq.~(\ref{eq:bloch_like_state}) is\cite{Sundaram_Niu_1999}
\begin{equation}
    L = -\mathcal{E} +\dot{\bm{r}} \vdot \pqty{ \bm{k} - e\bm{A}} 
    + \dot{\bm{r}} \vdot \bm{\mathcal{A}}_r + \dot{\bm{k}} \vdot \bm{\mathcal{A}}_k.
    \label{eq:lagrangian}
\end{equation}
Strictly speaking, the energy $\mathcal{E}$ includes gradient corrections and the wave packet Zeeman energy. However, the former produces effects that are parametrically small in $V/t$, while the latter vanishes in our model Hamiltonian, so these terms will be ignored; that is, we take $\mathcal{E}$ as the band energy. The Berry connections are
\begin{align}
    \bm{\mathcal{A}}_r &= \mel{c}{i \grad_{\bm{r}}}{ c} \\
    \bm{\mathcal{A}}_{k} &= \mel{c}{i \grad_{\bm{k}}}{ c}.
\end{align}

The EOM are
\begin{align}
\dot{\bm{r}} &= 
\grad_{\bm{k}} \mathcal{E} -
\qty( \Omega_{kr} \dot{\bm{r}} + \Omega_{kk} \dot{\bm{k}})   \\ 
\dot{\bm{k}} &=  -e\dot{\bm{r}} \cp \bm{B} 
 +\qty( \Omega_{r r} \dot{\bm{r}} + \Omega_{r k}  \dot{\bm{k}}) . 
\end{align}
where the Berry curvature tensors are, for example, 
\begin{equation}
    (\Omega_{k r})_{ab} = \pdv{(\mathcal{A}_r)_b}{k_a} - \pdv{(\mathcal{A}_k)_b}{r_a}.
\end{equation}
Since the energy $\mathcal{E}(\bm{k})$ is conserved, the $k$-space orbits coincide with the Fermi surface as in the absence of CDW disorder. 

\subsection{Berry curvatures}
Start with the eigenstate
\begin{equation}
c = \begin{pmatrix}
e^{-i\phi(\bm{r})/2} \sqrt{\frac{1}{2}
(1+\frac{\Delta \mathcal{E}_0}{\Delta \mathcal{E}})} \\
e^{i\phi(\bm{r})/2} \sqrt{\frac{1}{2}(1-\frac{\Delta \mathcal{E}_0}{\Delta \mathcal{E}})}
\label{eq:eigenstates}
\end{pmatrix}
\end{equation}
for the upper band of $\mathcal{H}$, 
where 
\begin{equation}
    \Delta \mathcal{E}_0(\bm{k}) = \mathcal{E}_0(\bm{k}) - \mathcal{E}_0(\bm{k} + \bm{Q})
\end{equation}
and $\Delta \mathcal{E}$ is the difference in energy between the upper and lower bands.
Then the Berry curvatures may be directly evaluated:
\begin{align}
    \Omega_{rr} &= 0 \\
    \Omega_{kk} &= 0 \\
    (\Omega_{kr})_{ab} &= \frac{1}{2} \bqty{ \pdv{k_a}(\frac{\Delta \mathcal{E}_0}{\Delta \mathcal{E}})
    }\pdv{\phi}{r_b}.
\end{align}
This simplifies in the strict $V/t \to 0$ limit, where
\begin{equation}
    (\Omega_{kr})_{ab} = \delta(k_x-\bar{k}_x)\bqty{ \delta_{ax} \pdv{\phi}{r_b}},
\end{equation}
and
\begin{equation}
    \bar{k}_x = \frac{\pi}{a} - \frac{Q}{2}
\end{equation}
is the vertical line $\Delta \mathcal{E}_0 = 0$. Small but finite $V/t$  broadens the $\delta$-function slightly, rounding out singularities or discontinuities that appear in the strict $V/t \to 0$ limit.

\subsection{Effect on classical trajectories}
In the presence of disorder, the $k$-space orbit still coincides with the Fermi surface, but Eq.~(\ref{eq:EOM_rotate}) is replaced by
\begin{equation}
    \dot{r}_a = \frac{1}{eB} \epsilon_{ab} \pqty{  \dot{k}_b +(\Omega_{kr})_{cb} \dot{k}_c  }
\end{equation}
which reduces in the $V/t \to 0$ limit to
\begin{equation}
    \dot{r}_a = \frac{1}{eB} \epsilon_{ab} \pqty{  \dot{k}_b +\pdv{\phi}{r_b}\dv{t} \theta(k_x-\bar{k}_x)}. \label{eq:EOM_disorder}
\end{equation}
This equation describes the displacement of the scattering points in the case where the CDW phase varies smoothly throughout. As a simple example, consider $\phi = (\delta Q) x$, $\delta Q \ll Q$. This describes a \emph{uniform} CDW with wavevector $Q + \delta Q$, so we expect that the top leg of the real-space orbit is displaced upward by $\delta Q/eB$ relative to constant $\phi$. Integrating the equation above, this is indeed the case.

Using Eq.~(\ref{eq:EOM_disorder}), we also find that after a single period, the guiding center drifts by
\begin{equation}
\Delta R_a = -\frac{1}{eB} \epsilon_{ab} \pdv{\gamma}{R_b}.
\end{equation}
This is perpendicular to $\pdv*{\gamma}{R_a}$, so, on average, $\gamma$ is conserved and $\bm{R}$ follows lines of constant $\gamma$. This average behavior may also be obtained by regarding the local Landau level energy
\begin{equation}
E_n(\bm{R}) =  - \omega_c \frac{\gamma(\bm{R})}{2\pi} + \text{constant} 
\end{equation}
as a classical Hamiltonian, where $R_x$ and $R_y$ satisfy the Poisson bracket ${\{ R_x, R_y \} = 1/eB}$.

\subsection{Berry phase}
In addition to possibly perturbing the classical trajectories, the Berry curvature gives to a Berry phase
\begin{equation}
    \gamma = \int \bm{\mathcal{A}}_r \vdot d \bm{r}
    + \bm{\mathcal{A}}_k \vdot d \bm{k}.
\end{equation}
Using the generalized stokes theorem, this can be written as:
\begin{equation}
    \gamma =  \int_{R} d^{2}r \hspace{2 pt} eB \Tr(\Omega_{kr}),
\end{equation}
where $R$ is the interior of the real-space orbit, and the argument $\bm{k}$ of $\Omega_{kr}$ is expressed in terms $\bm{r}$ using Eq.~(\ref{eq:guiding}). Written in this way, we may think of the electron as feeling a spatially fluctuating magnetic field ${b = eB\Tr(\Omega_{kr}) }$ that depends on the guiding center of the orbit under consideration, but not on its energy. In the strict $V/t \to 0$ limit
\begin{equation}
    b = \delta(y-\bar{y}) \pdv{\phi}{x}
\end{equation}
where $\bar{y}$ is the horizontal reflection axis of the orbit. In this limit, we directly obtain $\gamma = \Delta \phi$.

In this loose analogy, the case of an orbit crossing a DC corresponds to inserting a flux tube in the center of the orbit. As in the Aharanov-Bohm effect, this leaves the semiclassical dynamics unaffected, but modifies the action and the energy spectrum.

\section{Berry phase across scattering point \label{app:berry_avoided_crossing}}

Here we evaluate the Berry phase picked up after passing through a scattering point. Let us consider the interval $I_R = [t_i, t_f]$ in Sec.~(\ref{sec:berry}). To reiterate, the boundary conditions are $c(t_i)  \approx(1,0)$, $c(t_f)  \approx(0,1)$. 
The simplest eigenstate for the upper band of $\mathcal{H}$ in the vicinity of the scattering point is obtained by substituting $\phi(\bm{r}) \to \phi(\bm{r}_R)$ in Eq.~\ref{eq:eigenstates}. This has zero Berry connection, but the wrong boundary condition: ${c(t_i) \approx (e^{-i\phi(\bm{r}_R)/2},0)}$ and $c(t_f) \approx (0,e^{i\phi(\bm{r}_R)/2})$ at the end. To fix this, multiply by $e^{-i\alpha(t)}$ where $\alpha$ is any function satisfying $\alpha(t_i) = - \phi(\bm{r}_R)/2$ and $\alpha(t_f) = \phi(\bm{r}_R)/2$ at late times. This gauge transformation then changes the Berry phase to $\alpha(t_f) -\alpha(t_i) = \phi(\bm{r}_R)$.

\section{Perturbed action and open orbits \label{app:action_perturbation}}
In the main text, we assumed that we could neglect changes  in  the semiclassical trajectory produced by a spatially varying CDW phase, and thus focused entirely on the change in the Berry phase. This assumption is trivially justified in the case of an array of sharp DCs, where (in the appropriate limit of a weak CDW potential) the semiclassical trajectory is unaffected by disorder. 

In the case where the CDW phase varies smoothly, however, the trajectories are slightly modified: the scattering points are displaced by a distance of order $l^2_B \pdv*{\phi}{r}$, where $l_B = \sqrt{\hbar/eB}$ is the magnetic length. Moreover, the modified trajectories drift, so, strictly speaking, there is no closed orbit to quantize. Nevertheless, the effect of these modifications on the quantized energy levels is negligible, as we now demonstrate.

Firstly, in order to start with closed orbits we define a slightly modified reference Hamiltonian:
\begin{equation}
\widetilde{H} = H +e\bm{E} \vdot \bm{r},
\end{equation}
where $\bm{E}$ is chosen so that near some point $\bm{r}_0$, the resulting $E\times B$ drift cancels the local drift due to CDW disorder. This requires 
\begin{equation}
E \sim  \pqty{ \omega_c  l^2_B \pdv{\phi}{r} } B .
\end{equation}
Since the family of orbits of $\widetilde{H}$ with guiding centers near $\bm{r}_0$ is closed, we can obtain the corresponding energy levels semiclassically. 

Adding an electric field to Eq.~(\ref{eq:lagrangian}), the classical wave packet Lagrangian associated with $\widetilde{H}$ is
\begin{equation}
    \widetilde{L} = -\mathcal{E} +\dot{\bm{r}} \vdot \pqty{ \bm{k} - e\bm{A}} 
    + \bqty{     -e \bm{E} \vdot \bm{r}
    + \dot{\bm{r}} \vdot \bm{\mathcal{A}}_r + 
    \dot{\bm{k}} \vdot \bm{\mathcal{A}}_k}.
\end{equation}
Assume we have picked coordinates so that ${\bm{r}_0 = 0}$ (that is, the electric field is a small perturbation to the orbits under consideration). Then the terms in the brackets may be accounted for using
semiclassical perturbation theory.  
Specifically, if a Lagrangian is perturbed, $ {L \to L + \delta L}$,
then the classical orbits are perturbed
such that the total change in action is
\begin{equation}
\delta S = \int_{0}^{T} dt \hspace{ 2 pt} \delta L.
\end{equation}
A derivation for the case without Berry phases can be found in Ref.~\onlinecite{Bohigas_et_al_1995}. For the problem at hand, 
\begin{equation}
\delta S = -e \bm{E}  \vdot \ev{\bm{r}} T  + \gamma \label{eq:perturbed_action}
\end{equation}
where angled brackets denote the time average of $\bm{r}$.

This gives the obvious shift in the quantization condition for $\widetilde{H}$.  Now, however, the energies of the  corresponding states in $H$ must be corrected perturbatively in the strength of the electric field.  To first order, 
its expectation value in an eigenstate of $\widetilde{H}$ yields $-e \bm{E} \vdot \ev{\bm{r}}$, canceling the shift in the energy levels due to the first term in Eq.~(\ref{eq:perturbed_action}). This leaves behind the effect of the Berry phase $\gamma$, plus corrections $\delta E$ which appear in higher order perturbation theory. Since the energy gap is $\hbar \omega_c$ and the matrix elements to higher/lower Landau levels scale as $e  E l_B $,
\begin{align}
\frac{\delta E}{\hbar \omega_c} &= \pqty{ \frac{e E l_B} {\hbar \omega_c}}^2 \\
&= \pqty{l_B \pdv{\phi}{r}}^2 \ll 1,
\end{align}
so that, under the assumption that the phase disorder is short-range correlated, to excellent approximation the energy spectrum can be obtained by simply shifting each orbit's action by $\gamma$.

\bibliography{bibliography.bib}

\end{document}